\documentclass[aps,twocolumn]{revtex4}
\pdfoutput=1

\usepackage{graphicx}
\usepackage{times}
\usepackage{amsmath}
\usepackage{amssymb}

\begin{document}
\title{Selected-fit versus induced-fit protein binding: \\Kinetic differences and mutational analysis}

\author{\hspace*{-1cm} Thomas R.\ Weikl and Carola von Deuster\\[0.2cm]
\hspace*{-1cm} \small Max Planck Institute of Colloids and Interfaces, Department of Theory 
\\[-0.1cm] 
\hspace*{-1cm} \small and Bio-Systems, Science Park Golm, 14424 Potsdam, Germany}

\begin{abstract}
The binding of a ligand molecule to a protein is often accompanied by conformational changes of the protein. A central question is whether the ligand induces the conformational change (induced-fit), or rather selects and stabilizes a complementary conformation from a pre-existing equilibrium of ground and excited states of the protein (selected-fit). We consider here the binding kinetics in a simple four-state model of ligand-protein binding. In this model, the protein has two conformations, which can both bind the ligand. The first conformation is the ground state of the protein when the ligand is off, and the second conformation is the ground state when the ligand is bound. The induced-fit mechanism corresponds to ligand binding in the unbound ground state, and the selected-fit mechanism to ligand binding in the excited state.  We find a simple, characteristic difference between the on- and off-rates in the two mechanisms if the conformational relaxation into the ground states is fast. In the case of selected-fit binding, the on-rate depends on the conformational equilibrium constant, while the off-rate is independent. In the case of induced-fit binding, in contrast, the off-rate depends on the conformational equilibrium, while the on-rate is independent. Whether a protein binds a ligand via selected-fit or induced-fit thus may be revealed by mutations far from the protein's binding pocket, or other ``perturbations'' that only affect the conformational equilibrium. In the case of selected-fit, such mutations will only change the on-rate, and in the case of induced-fit, only the off-rate.
\end{abstract}

\maketitle

\section*{INTRODUCTION}

\begin{figure}[t]
\resizebox{0.78\linewidth}{!}{\includegraphics{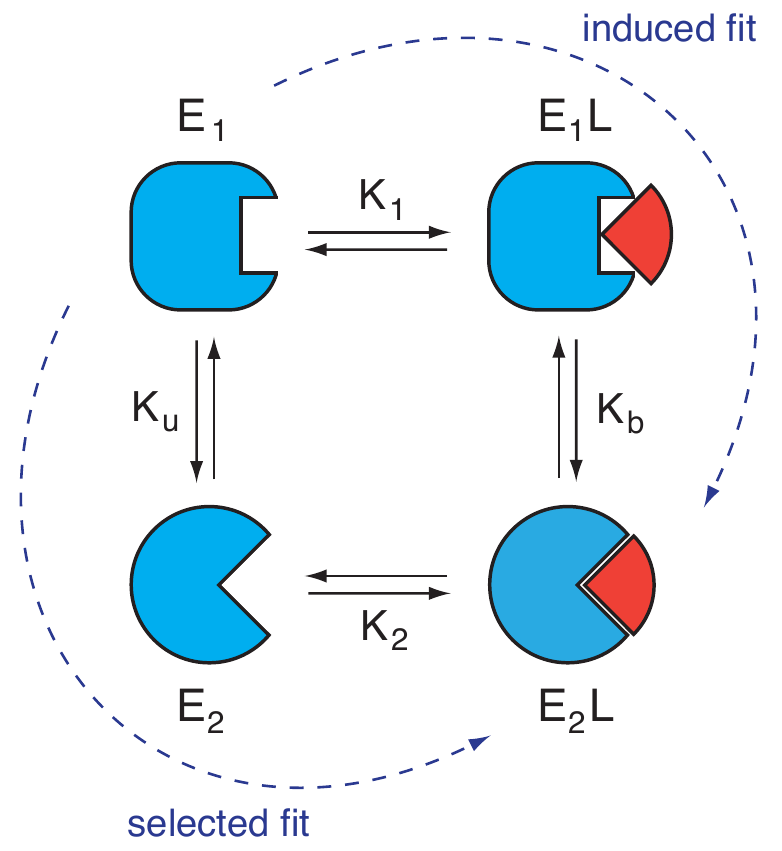}}
\vspace{-0.2cm}
\caption{Simple four-state model of protein-ligand binding. Without the ligand $L$, the conformation $E_1$ of the protein, or enzyme, is the ground-state conformation, and $E_2$ is the excited-state conformation. When the ligand is bound, $E_2L$ is the ground state, and $E_1L$ the excited state. On the selected-fit  route, the ligand binds the protein in the excited-state conformation $E_2$, and on the induced-fit binding route, in the ground-state conformation $E_1$.}
\label{figure_cartoon}
\end{figure}

The function of a protein often requires dynamic changes between different three-dimensional conformations. 
Conformational changes that occur when a protein binds a ligand molecule are well documented by experimentally determined structures of proteins in the unbound state and the ligand-bound state \cite{Gerstein98,Goh04}. In 1958, Koshland \cite{Koshland58} suggested that the binding of the ligand may induce the conformational change of the protein. More recently, Tsai {\it et al.} \cite{Tsai99,Tsai01} suggested an alternative mechanism in which the ligand selects and stabilizes a complementary protein conformation from an equilibrium of low-energy and higher-energy conformations. This selected-fit, or conformational selection mechanism is based on the energy-landscape picture of protein folding \cite{Dill97,Bryngelson95}.

How can we distinguish whether a protein binds its ligand in an induced-fit or selected-fit mechanism? Recent single-molecule fluorescence \cite{Lu98,Ha99,Antikainen05,Min05,Michalet06,Smiley06} and NMR relaxation experiments \cite{Eisenmesser02,Wang04,Boehr06a,Boehr06b,Henzler07} provide dynamic insights into conformational transitions. NMR relaxation experiments of the enzyme DHFR, which binds two ligands, reveal excited-state conformations that resemble ground-state conformations of adjacent ligand-bound or ligand-free states on the catalytic cycle \cite{Boehr06a}. These experiments provide evidence for a pre-existing equilibrium of conformations that is shifted by the binding or unbinding of a ligand, as suggested in the selected-fit mechanism of protein binding \cite{Tsai99,Tsai01}. For the binding of myosin to actin, in contrast, an induced-fit mechanism has been proposed \cite{Goh04} based on cryo-microscopic images of the actin-myosin complex \cite{Holmes03}.

We consider here the selected-fit and induced-fit binding kinetics in a simple four-state model of protein-ligand binding (see fig.~\ref{figure_cartoon}). In this model,  the protein, or enzyme, has two dominant conformations $E_1$ and $E_2$. The conformation $E_1$ is the ground-state conformation in the unbound state of the protein, while $E_2$ is the ground-state conformation in the ligand-bound state.  Two routes connect the unbound ground state $E_1$ and bound ground state $E_2L$. On the induced-fit route $E_1 \to E_1 L \to E_2 L$, the protein first binds the ligand in conformation $E_1$, which causes the transition into conformation $E_2$. On the selected-fit route $E_1 \to E_2 \to E_2 L$, the protein binds the ligand in the higher-energy conformation $E_2$. 

We find a characteristic difference between the selected-fit and induced-fit binding kinetics. If the conformational relaxation into the ground state is fast, the selected-fit on-rate depends on the equilibrium constant of the conformations $E_1$ and $E_2$, while the selected-fit off-rate is independent of the conformational equilibrium. The induced-fit on-rate, in contrast, is independent of the conformational equilibrium between $E_1 L$ and $E_2 L$, whereas the induced-fit off-rate depends on this equilibrium. Mutations or other perturbations that shift the conformational equilibrium without affecting the shape or free energies of the binding site thus may help to identify whether a protein binds its ligand in a selected-fit or induced-fit mechanism.

\section*{RESULTS}

%
\subsection*{Equilibrium behavior}

The four states of our model are connected by four transitions with equilibrium constants $K_1$, $K_2$, $K_u$, and $K_b$ (see fig.~\ref{figure_cartoon}). In the model, the constants $K_u$ and $K_b$ for the conformational equilibrium in the unbound and the bound state obey 
\begin{equation}
K_u = \frac{[E_2]}{[E_1]}\ll 1 \;, \;\;\; K_b = \frac{[E_2L]}{[E_1L]}\gg 1
\end{equation}
 since $E_1$ is the ground-state conformation in the unbound state of the protein, and $E_2 L$ is the ground state when the ligand is bound. We assume here that the excited-state conformations are significantly higher in free energy than the ground states. The binding equilibrium of the two conformations is governed by the constants
\begin{equation}
K_1 = \frac{[E_1 L]}{[E_1][\rm L]} \;, \;\;\; K_2 = \frac{[E_2 L]}{[E_2][L]}
\end{equation}

From the two equalities $[E_2 L] = K_2 [E_2] [L] = K_2 K_u [E_1] [L]$ and $[E_2 L] = K_b [E_1 L] = K_b K_1 [E_1] [L]$ that follow directly from these definitions, we obtain the general relation
\begin{equation}
K_u K_2= K_b K_1  \label{equilibrium_relation}
\end{equation}
between the four equilibrium constants. Thus, only three of the equilibrium constants are independent. 

The selected-fit route $E_1 \to E_2 \to E_2 L$ in our model dominates over the induced-fit route $E_1 \to E_1 L \to E_2 L$ if the concentration $[E_2]$ of the selected-fit intermediate is larger than the concentration $[E_1 L]$ of the induced-fit intermediate. Since $[E_2] = K_u [E_1]$ and  $[E_1 L] = K_1 [E_1] [L]$, the selected-fit mechanism thus is dominant for small ligand concentrations 
\begin{equation}
[L]< K_u/K_1 \label{selected_fit_dominates}
\end{equation}
while the induced-fit mechanism is dominant for large ligand concentrations $[L]> K_u/K_1$.

\subsection*{Selected-fit binding kinetics}

\begin{figure}[b]
\vspace*{0.5cm}
\begin{center}
\resizebox{0.95\linewidth}{!}{\includegraphics{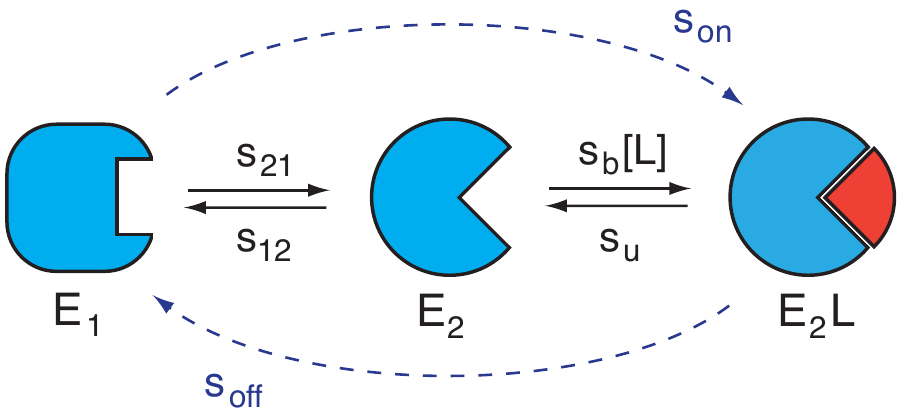}}
\end{center}
\vspace*{-0.2cm}
\caption{Binding kinetics along the selected-fit route of our model (see fig.~\ref{figure_cartoon}). Here, $s_{21}$ and $s_{12}$ are the rates for the conformational transitions in the unbound state, $s_b$ is the binding rate of conformation $E_2$ per mole ligand, and $s_u$ the unbinding rate. Since $E_1$ is the ground state and $E_2$ the excited state, we have $K_u = s_{21}/s_{12} \ll 1$. If the conformational transition rate $s_{12}$ into the ground state is much larger than the binding and unbinding rates $s_b[L]$ and $s_u$, the on-rate along the selected-fit route is approximately $s_\text{on}\approx K_u s_b$, and the off-rate is $s_\text{off} \approx s_u$ (see eqs.~(\ref{son}) and (\ref{soff})).}
\label{figure_selected_fit}
\end{figure}

We first consider the binding kinetics along the selected-fit route of our model (see figs.~\ref{figure_cartoon} and \ref{figure_selected_fit}). The selected-fit binding rate is the dominant relaxation rate of the process 
\begin{equation}
E_1 \;
\begin{matrix} 
\text{\footnotesize $s_{21}$}\\[-0.05cm]
\text{\Large $\rightleftharpoons$}\\[-0.2cm]
\text{\footnotesize $s_{12}$} \\[0.1cm]
\end{matrix}
 \; E_2 \;
\begin{matrix} 
\text{\footnotesize $s_b[L]$}\\[-0.1cm]
\text{\Large $\longrightarrow$}\\[-0.1cm]
~ 
\end{matrix}
\; E_2L
\label{selectedfit_on}
\end{equation}
%
from $E_1$ to $E_2L$, with $E_2L$ as an ``absorbing state" without backflow into $E_2$. Here, $s_{21}$ is the transition rate from state $E_1$ to $E_2$, $s_{12}$ is the rate for the reverse transition, and $s_b$ is the binding rate per mole ligand in state $E_2$.
In the appendix, we calculate the exact relaxation rates for a process of the form (\ref{selectedfit_on}). Since $E_2$ is the excited state, we have $s_{21}/s_{12} = K_u \ll 1$, and the on-rate of the selected-fit process (\ref{selectedfit_on}) is approximately given by 
\begin{equation}
s_\text{on} [L] \approx \frac{s_\text{21}s_b[L]}{s_\text{12}+s_b[L]}  
\label{son_one}
\end{equation}
(see eq.~(\ref{rate_one}) in the Appendix). For small ligand concentrations $[L]$, or fast conformational relaxation $s_\text{12}$ into the ground-state conformation $E_1$, we have $s_b[L] \ll s_\text{12}$, and the selected-fit on-rate per mole ligand is 
\begin{equation}
\fbox{$\displaystyle s_\text{on} \approx \frac{s_\text{21}s_b}{s_\text{12}} = K_u s_b $} 
\label{son}
\end{equation}
The selected-fit on-rate (\ref{son}) just depends on the equilibrium constant $K_u$ of the conformations $E_1$ and $E_2$ in the unbound state, and the binding rate $s_b$ of the conformation $E_2$. Since the equilibrium probability $P(E_2)=[E_2]/([E_1]+[E_2])$ of conformation $E_2$ is approximately $P(E_2)\approx [E_2]/[E_1]=K_u$ for $[E_2]\ll [E_1]$, the selected-fit on-rate (\ref{son}) can also be directly understood as the product of the probability that the protein is in conformation $E_2$ and the binding rate $s_b$ of this conformation. 

The selected-fit off-rate is the dominant relaxation rate of the process 
\begin{equation}
E_2 L \;
\begin{matrix} 
\text{\footnotesize $s_u$}\\[-0.05cm]
\text{\Large $\rightleftharpoons$}\\[-0.2cm]
\text{\footnotesize $s_b [L]$} \\[0.1cm]
\end{matrix}
 \; E_2 \;
\begin{matrix} 
\text{\footnotesize $s_{12}$}\\[-0.1cm]
\text{\Large $\longrightarrow$}\\[-0.1cm]
~ 
\end{matrix}
\; E_1
\label{selectedfit_off}
\end{equation}
from $E_2 L$ to the $E_1$. This process follows the same general reaction scheme as the binding process (\ref{selectedfit_on}). A reasonable, simplifying assumption here is that the conformational relaxation process from the excited state $E_2$ to the ground state $E_1$ is significantly faster than the binding and unbinding process, which implies $s_{12} \gg  s_u$ and $s_{12} \gg  s_b [L]$. The selected-fit off-rate then simply is 
\begin{equation}
\fbox{$\displaystyle s_\text{off} \approx s_{u}$} 
\label{soff}
\end{equation}
(see eq.~(\ref{rate_two}) in the Appendix). The off-rate $s_\text{off}$ thus is independent of the conformational transition rates between $E_1$ and $E_2$. The off-rate is identical with the rate $s_u$ for the bottleneck step, the unbinding process from $E_2L$ to $E_2$. 

\subsection*{Induced-fit binding kinetics}

\begin{figure}
\begin{center}
\resizebox{0.95\linewidth}{!}{\includegraphics{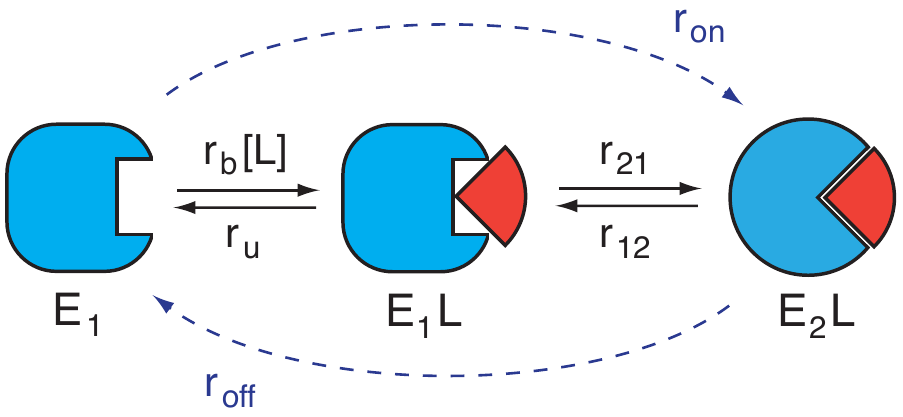}}
\end{center}
\vspace{-0.2cm}
\caption{Binding kinetics along the induced-fit route of our model (see fig.~\ref{figure_cartoon}). Here, $r_{12}$ and $r_{21}$ are the rates for the conformational transitions in the bound state, $r_b$ is the binding rate of conformation $E_1$ per mole ligand, and $r_u$ the unbinding rate. Because $E_2 L$ is the ground state, we have $K_b = r_{21}/r_{12}\gg 1$. For conformational transition rates $r_{21}$ into the bound ground state that are much larger than the binding and unbinding rates $r_b[L]$ and $r_u$, the on- and off-rates along the induced-fit route are approximately $r_\text{on}\approx r_b$ and $r_\text{off}\approx s_u/K_b$ (see eqs.~(\ref{ron}) and (\ref{roff})).}
\label{figure_induced_fit}
\end{figure}

The on-rate along the induced-fit binding route of our model (see figs.~\ref{figure_cartoon} and \ref{figure_induced_fit}) is the dominant relaxation rate of the process
\begin{equation}
E_1 \;
\begin{matrix} 
\text{\footnotesize $r_b[L]$}\\[-0.05cm]
\text{\Large $\rightleftharpoons$}\\[-0.2cm]
\text{\footnotesize $r_u$} \\[0.1cm]
\end{matrix}
 \; E_1 L \;
\begin{matrix} 
\text{\footnotesize $r_{21}$}\\[-0.1cm]
\text{\Large $\longrightarrow$}\\[-0.1cm]
~ 
\end{matrix}
\; E_2L
\label{inducedfit_on}
\end{equation}
Here, $r_b$ is the binding rate of conformation $E_1$ per mole ligand, $r_u$ is the unbinding rate, and $r_{21}$ is the rate for the conformational transition into the bound ground state $E_2 L$. The induced-fit binding process (\ref{inducedfit_on}) is similar to the selected-fit unbinding process (\ref{selectedfit_off}). As before, we assume that the conformational transition into the ground state is much faster than the binding and unbinding processes, i.e.~we assume $r_{21}\gg r_b[L]$ and $r_{21}\gg r_u$. The induced-fit on-rate per mole ligand then is (see Appendix)
\begin{equation}
\fbox{$\displaystyle r_\text{on} \approx r_{b}$} 
\label{ron}
\end{equation}
%

The induced-fit unbinding rate is the dominant relaxation rate of the process
\begin{equation}
E_2 L \;
\begin{matrix} 
\text{\footnotesize $r_{12}$}\\[-0.05cm]
\text{\Large $\rightleftharpoons$}\\[-0.2cm]
\text{\footnotesize $r_{21}$} \\[0.1cm]
\end{matrix}
 \; E_1 L \;
\begin{matrix} 
\text{\footnotesize $r_u$}\\[-0.1cm]
\text{\Large $\longrightarrow$}\\[-0.1cm]
~ 
\end{matrix}
\; E_1 
\label{inducedfit_off}
\end{equation}
which is similar to the selected-fit binding process (\ref{selectedfit_on}). Since $E_2L$ is the ground-state conformation in the bound state of the protein, we have $K_b=r_{21}/r_{12}\gg 1$. The dominant relaxation rate of (\ref{inducedfit_off}) then is 
\begin{equation}
 r_\text{off} \approx \frac{r_{12}r_u}{r_{21}+r_u} 
\end{equation}
(see eq.~(\ref{rate_one}) in the Appendix). For fast conformational relaxation into the bound ground state with rate $r_{21}\gg r_u$,
the induced-fit off-rate is approximately
\begin{equation}
\fbox{$\displaystyle r_\text{off} \approx \frac{r_u}{K_b}$} 
\label{roff}
\end{equation}
The off-rate $r_u/K_b$ can again be understood as the product of the unbinding rate $r_u$ in the excited state $E_1L$ and the equilibrium probability $P(E_1L) = [E_1L]/([E_1L]+[E_2L])$ of this state, which is approximately $P(E_1L) \approx [E_1L]/[E_2L] = 1/K_b$. 

\subsection*{Mutational analysis of the binding kinetics}

A mutational analysis of the binding kinetics may reveal the characteristic differences between the selected-fit and induced-fit on-rates (\ref{son}) and (\ref{ron}) and off-rates (\ref{soff}) and (\ref{roff}). Of particular interest here are mutations of residues far away from the ligand-binding site that affect the rate constants for transitions between the conformations, but not the binding rates $s_b$ and $r_b$ and unbinding rates $s_u$ and $r_u$ of the two conformations. The mutations will change the free-energy differences $\Delta G_u = G(E_2) - G(E_1)$ and $\Delta G_b = G(E_2L) - G(E_1L)$ between the two conformations in the unbound and the bound state. In terms of these free-energy differences, the equilibrium constants $K_u$ and $K_b$ for the conformational transitions are
\begin{equation}
K_\text{u} = RT \exp(-\Delta G_u/RT) \;\; \text{and} \;\; K_\text{b} = RT \exp(-\Delta G_b/RT)
\end{equation}
From eq.~(\ref{son}), we find that the ratio of the selected-fit on-rates for wildtype and mutant 
\begin{equation}
\fbox{$\displaystyle \frac{s^\prime_\text{on}}{s_\text{on}}=\frac{K^\prime_u}{K_u} = \exp(-\Delta\Delta G_u/RT)$}
\label{selectedfit_mut}
\end{equation}
depends on the mutation-induced shift $\Delta \Delta G_u = \Delta G^\prime_u - \Delta G_u$  of the free energy difference between the conformations in the unbound state. The prime here indicates the conformational free-energy difference for the mutant. From eq.~(\ref{roff}), we obtain the relation
\begin{equation}
\fbox{$\displaystyle \frac{r^\prime_\text{off}}{r_\text{off}}=\frac{K_b}{K^\prime_b} = \exp(\Delta\Delta G_b/RT)$}
\label{inducedfit_mut}
\end{equation}
between the induced-fit off-rates of wildtype and mutant. Here, $\Delta \Delta G_b = \Delta G^\prime_b - \Delta G_b$ is the mutation-induced shift of the conformational free-energy difference in the bound state. The selected-fit off-rate (\ref{soff}) and induced-fit on-rate (\ref{ron}), in contrast, are the same for wildtype and mutant if the mutations do not affect the binding and unbinding rates of the two conformations.

\section*{DISCUSSION AND CONCLUSIONS}

Empirical methods to calculate mutation-induced stability changes of proteins have been investigated intensively in the past years \cite{Guerois02,Schymkowitz05,Gromiha99b,Gilis00,Zhou02b,Capriotti05,Cheng06,Huang07b,Parthiban07,Bueno07}. These methods can be used to calculate the changes $\Delta\Delta G_u$ of the conformational free-energy difference in the unbound state, provided that experimental protein structures are available both for the unbound and ligand-bound state of the protein. The experimental structure of the protein in the ligand-bound state then corresponds to the excited-state conformation in the unbound state of our model. Calculations of mutation-induced changes $\Delta\Delta G_b$ of conformational free-energy differences in the bound state are more difficult and require a construction of a ligand-bound excited-state conformation by docking the ligand to the experimental structure of the unbound protein.

In combination with eqs.~(\ref{selectedfit_mut}) and (\ref{inducedfit_mut}) of our model, calculated free-energy changes may help to identify selected-fit or induced-fit mechanisms of protein binding. A promising candidate is the protein DHFR, which has been studied extensively as a model enzyme to understand the relations between conformational dynamics, binding, and function.  The enzymatic mechanism and conformational changes of DHFR have been investigated by mutational analyses \cite{Cameron97,Miller98d,Miller98c,Miller01,Wang06}, NMR relaxation \cite{Boehr06a,Boehr06b,McElheny05}, single-molecule fluorescence experiments \cite{Antikainen05}, and simulation \cite{Agarwal02,Rod03,Chen07}. The strong effect of point mutations far from the ligand binding site on catalytic rates and binding rates suggests large conformational changes during the catalytic cycle \cite{Benkovic08,Cameron97,Miller98c,Miller01,Wang06}, in agreement with experimental structures of DHFR in the unbound and several ligand-bound states \cite{Sawaya97}. Mutations far from the binding sites of DHFR affect the binding kinetics of the two ligands \cite{Miller98d,Miller98c,Miller01}, and NMR relaxation experiments point towards a selected-fit mechanism of ligand binding \cite{Boehr06a}. However, the mutations seem to affect both the binding and the unbinding rates \cite{Miller98d,Miller98c,Miller01}. This precludes a direct comparison with our model, either because the conformational dynamics is more complex, or because the mutations have an indirect effect on the binding free energies. 

We have considered here a simple four-state model of protein-ligand binding. The kinetic and mutational analysis of this model is partly inspired by previous models of the protein folding kinetics \cite{Merlo05,Weikl08}. We have found characteristic kinetic differences between the selected-fit and induced-fit binding routes of this model, which should be observable also in more detailed models of conformational changes and binding. While the timescale of the relevant conformational transitions are in general inaccessible to standard molecular dynamics simulations with atomistic protein models \cite{Karplus02}, conformational selection during binding has been recently studied by combining atomistic simulations and docking \cite{Wong08,Wong05,Frembgen06}. Large conformational transitions are also intensively studied by normal mode analysis \cite{Ma05,Case94}, in particular with coarse-grained elastic network models of proteins  \cite{Bahar05,Yang07,Tirion96,Bahar97,Micheletti04,Zheng07}.

\section*{APPENDIX: RELAXATION RATES}

%
\subsection*{General solution}

The binding and unbinding processes of our model have a general reaction scheme of the form 
\begin{equation}
A \rightleftharpoons B \rightarrow C    \label{scheme1}
\end{equation}
For the selected-fit or induced-fit binding process, state $A$ corresponds to $E_1$, and state $C$ to $E_2 L$.  For the unbinding processes, $A$ corresponds to  $E_2 L$, and $C$ to $E_1$. We are interested in the dominant, slowest relaxation rate into state $C$. Therefore, the state $C$ is an ``absorbing'' state, i.e.~the rate from $C$ back to $B$ is 0. The probability evolution $P_A(t)$, $P_B(t)$, and $P_C(t)$ of the three states is then governed by the set of master equations
\begin{eqnarray}
\frac{d P_A(t)}{dt} &=& k_{AB} P_B(t) - k_{BA} P_A(t)  \label{me_one}\\
\frac{d P_B(t)}{dt} &=& k_{BA} P_A(t) - (k_{AB} + k_{CB })P_B(t) \\
\frac{d P_C(t)}{dt} &=& k_{CB} P_B(t) \label{me_three}
\end{eqnarray}
where $k_{ij}$ denotes the rate from state $j$ to state $i$. The three equations can be written in the matrix form
\begin{equation}
\frac{d \boldsymbol{P}(t)}{dt} = - \boldsymbol{W} \boldsymbol{P}(t)  \label{matrix_equation}
\end{equation}
with $\boldsymbol{P}(t)=\left(P_A(t),P_B(t),P_C(t)\right)$ and 
\begin{equation}
\boldsymbol{W} = 
\begin{pmatrix} 
k_{BA} & - k_{AB} & 0\\
- k_{BA} & k_{AB} + k_{CB} & 0\\
0 & -k_{CB} & 0 
\end{pmatrix}
\end{equation}
The general solution of (\ref{matrix_equation}) has the form \cite{Kampen92}
\begin{equation}
\boldsymbol{P}(t) = c_o \boldsymbol{Y}_o + c_1 \boldsymbol{Y}_1 e^{-\lambda_1 t} + c_2 \boldsymbol{Y}_1 e^{-\lambda_2 t}
\label{gensol}
\end{equation}
where $\boldsymbol{Y}_o$,  $\boldsymbol{Y}_1$, and $\boldsymbol{Y}_2$ are the three eigenvectors of the matrix $\boldsymbol{W}$, and $\lambda_1$ and $\lambda_2$ are the two positive eigenvalues of the eigenvectors $\boldsymbol{Y}_1$ and $\boldsymbol{Y}_2$. The coefficients $c_o$, $c_1$, and $c_2$ depend on the initial conditions at time $t=0$. The eigenvalue of the eigenvector $\boldsymbol{Y}_o$ is 0, which ensures that $\boldsymbol{P}(t)$ relaxes towards a finite equilibrium probability distribution $\boldsymbol{P}(t\to \infty) = c_o \boldsymbol{Y}_o$ for $t \to \infty$. The two nonzero eigenvalues are
\begin{eqnarray}
\lambda_{1/2}=\frac{1}{2} \bigg(k_{AB} + k_{BA} + k_{CB} \hspace*{4cm}\nonumber
\\\hspace*{1cm} \pm \sqrt{k_{AB}^2 + (k_{BA} - k_{CB})^2 + 2 k_{AB} (k_{BA} + k_{CB})}
   \bigg) \nonumber
\end{eqnarray}
These eigenvalues, and their eigenvectors $\boldsymbol{Y}_1$ and $\boldsymbol{Y}_2$, capture the time-dependent relaxation process into equilibrium (see eq.~(\ref{gensol})).

\subsection*{Selected-fit binding and induced-fit unbinding}

In selected-fit binding (\ref{selectedfit_on}), state $A$ corresponds to $E_1$, state $B$ to $E_2$, and $C$ to $E_2L$ (see fig.~\ref{figure_cartoon}). In induced-fit unbinding (\ref{inducedfit_off}), $A$ corresponds to $E_2L$, $B$ to $E_1L$ and $C$ to $E_1$. In our model, we have $[E_2] \ll [E_1]$ and $[E_2L] \gg [E_1L]$ in equilibrium, which implies 
 $k_{BA} \ll k_{AB}$ in both cases. The nonzero eigenvalues then are approximately 
\begin{equation}
\lambda_1 \approx \frac{k_{BA}k_{CB}}{k_{AB}+k_{CB}}  \;\;\; \mbox{and} \;\;\; \lambda_2 \approx k_{AB} + k_{CB} \label{rate_one}
\end{equation}
For $k_{BA} \ll k_{AB}$, we have $\lambda_1 \ll \lambda_2$. The smaller rate $\lambda_1$ is the dominant relaxation rate for the initial conditions $\boldsymbol{P}(0)= (1,0,0)$ (only state $A$ populated) or $\boldsymbol{P}(0)= (1-k_{BA}/k_{AB},k_{BA}/k_{AB},0)$ (`pre-equilibration' of states $A$ and $B$). For both initial conditions, the probability evolution of state $C$ is approximately given by $P_C(t) \approx 1 - \exp[-\lambda_1 t]$. The dominant relaxation rate $\lambda_1$ of eq.~(\ref{rate_one}) can also be obtained from  eqs.~(\ref{me_one}) to (\ref{me_three}) in a steady-state approximation, i.e.~under the assumption that the variation $d P_B(t)/dt$ of the intermediate state $B$ is negligibly small.

\subsection*{Selected-fit unbinding and induced-fit binding}

In selected-fit unbinding (\ref{selectedfit_off}), the rate $k_{CB}$ corresponds to $s_{12}$, the rate from the excited state $E_2$ in the unbound ground  state $E_1$. In induced-fit binding (\ref{inducedfit_on}), $k_{CB}$ corresponds to $r_{21}$, the rate from $E_1 L$ in the bound ground state $E_2 L$. We assume here that the rates for these conformational transitions into the ground state are much larger than the binding and unbinding rates. This  implies $k_{AB}\ll k_{CB}$ and $k_{BA}\ll k_{CB}$. The nonzero eigenvalues then simply are
\begin{equation}
\lambda_1 \approx k_{BA} \;\;\; \mbox{and} \;\;\; \lambda_2 \approx k_{CB}
\label{rate_two}
\end{equation}
with the clearly smaller eigenvalue $\lambda_1$ as the dominant rate of the relaxation process from state $A$ to $C$.

\end{document}